\documentclass[preprint,12pt]{elsarticle}
\usepackage{graphicx}
\usepackage{amssymb}
\usepackage{lineno}
\usepackage{multicol}
\usepackage[dvipsnames]{xcolor}
\usepackage{geometry}
\usepackage{subcaption}

\begin{document}
\begin{frontmatter}
\title{Up-sampling of electron beam simulation particles with addition of shot-noise}
\author{P. Traczykowski$^{1,2,3}$, L.T. Campbell$^{1,2,3}$, B.W.J. M$^{\mathrm c}$Neil$^{1,2,*}$\corref{cor1} \\
              1 SUPA, Department of Physics, University of Strathclyde, Glasgow, G4 0NG\\
              2 Cockcroft Institute, Warrington, WA4 4AD, UK\\
              3 ASTeC, STFC Daresbury Laboratory, Warrington, WA4 4AD, UK}
\cortext[cor1]{Corresponding author email address: b.w.j.mcneil@strath.ac.uk}

\begin{abstract}
An algorithm and numerical code for the up-sampling of a system of particles, from a smaller to a larger number, is described. The method introduces a Poissonian `shot-noise' to the up-sampled distribution, typical of the noise statistics arising in a bunch of particles generated by a particle accelerator. The algorithm is applied to a 6-Dimensional phase-space distribution of relatively few simulation particles, representing an electron beam generated by particle accelerator modelling software, for subsequent injection into an Free Electron Laser (FEL) amplifier which  is used here to describe the model. A much larger number of particles is usually required to model the FEL lasing process than is required to model the electron beam accelerators that drive it. 
A numerical code developed from the algorithm was then used to generate electron bunches for injection into to an unaveraged 3D FEL simulation code, Puffin. Results show good qualitative and quantitative agreement with analytical theory. The program and user manual is available for download.
\end{abstract}

\end{frontmatter}

\section{Introduction}
Particle In Cell software usually runs faster with fewer simulation particles, with the optimum number determined by the degree of finesse required to model the physical processes. Sometimes it is necessary to increase the degree of detail being modelled in moving from one physical process to another. This may require an increase in the number of simulation particles in moving between simulation codes.
For example, when modelling an X-ray Free Electron Laser (FEL)~\cite{np}, the number of simulation particles used to model the acceleration stages of the FEL  often needs to be increased before simulation of the FEL itself. This is mainly due to the fine longitudinal electron bunching structures at the FEL radiation wavelength that need to be modelled. 
Such fine detail is very often not used when modelling the accelerator stages before the FEL as it would be computationally inefficient, for example when simulations are used to optimise these stages by scanning over different parameters.
The data set from the accelerator stages therefore often have a relatively sparse `{\it macroparticle}' distribution in phase space, each of which represents many electrons, and which do not model the Poissonian noise statistics of a real electron distribution. Here, a method is described which breaks up this sparse phase--space distribution of macroparticles into a greater number of `{\it microparticles}', each representing fewer electrons, to give a more dense phase--space distribution that is suitable for injection into an FEL simulation code such as Puffin~\cite{puffin,puffin2}, which is used here as the target code to demonstrate the methods used. Furthermore, the method ensures that the microparticle distribution has the correct Poissonian shot--noise statistics of a real electron beam~\cite{shotnoise}. This is necessary to simulate the spontaneous light generation which arises from the shot-noise and acts as the seed field from which shorter wavelength amplifier FELs start up in the Self Amplified Spontaneous Emission (SASE) mode of operation~\cite{np}. 

Of course, any fine phase-space structures that may develop in a real electron beam, but that are not captured by the macroparticle distribution from the accelerator/beam transport system simulations, would not be present in any up-sampled microparticle distribution. A smoothing of the macroparticle distribution is carried out before the microparticles are assigned: too small a smoothing function width and the individual macroparticles remain visible in the microparticle distribution; too wide and finer structures in the macroparticle distribution are smoothed out. The widths of these smoothing functions therefore limit the scales at which a real electron beam can be modelled by the final microparticle distribution, and need to be chosen carefully. It is clearly important to maintain the beam characteristics, such as current, emittance, energy, energy spread and spatial dimensions, as close to the original accelerator simulator generated macroparticle distribution as possible. However, given the wide range of possible particle distributions that may occur, it would be difficult to prescribe a general method to do this and it is left to the user to choose smoothing parameters appropriate for their requirements.

In Sections \ref{S2} to \ref{S5} the method of generating the microparticle distribution is described in more detail. Before a FEL simulation using Puffin, the microparticles should be initialised by distributing them longitudinally along the electron beam propagation direction, here the $z$--axis, with sufficient longitudinal density to allow modelling of variations that are smaller than the resonant radiation wavelength -- such variations give rise to coherent emission in the FEL~\cite{np}. In Puffin, the thickness of each longitudinal slice (or bin) in $z$ is therefore less than a resonant radiation wavelength. The longitudinal $z$--axis is first discretised into a series of transverse slices of sufficiently small width, each of which can contain a different charge (real number of electrons). The microparticles are then distributed in transverse phase space $(x,y)$ within each of these slices and the appropriate charge weighting and Poissonian shot-noise applied.  The initialisation slicing structure is then removed and the microparticles are free to interact with the radiation field and evolve in 6-Dimensional phase space without confinement to any sliced structure.

Other FEL simulation codes, which use the Slowly Varying Envelope Approximation (SVEA) and cannot therefore model sub-wavelength radiation evolution, such as GENESIS~\cite{genesis,genesis2}, allocate their initial microparticle distributions according to the charge in single or multiple wavelength longitudinal slices. GENESIS simulations therefore have fewer slices into which the microparticles are initially assigned. In order to maintain the correct shot-noise, microparticles remain in these slices during the FEL interaction. If, however, each microparticle represents a single electron, then the microparticles (electrons) can be re-distributed between slices as the FEL simulation progresses while maintaining the correct shot-noise statistics.

\section{Charge density distribution functions \label{S2}}
The method first uses the macroparticles of the accelerator modelling stage (each of which may have an integer number of electron charges) to create a discrete charge histogram of bin width $\Delta z_h$ along the longitudinal $z$-axis of beam propagation at a given time (e.g. on entering the FEL). 
A continuous longitudinal charge density function $f_z(z)$, proportional to the beam current, is then created by interpolating from this histogram. 
The beam is then discretised into a series of `slices' of width $\Delta z_s$ and charge $f_z(z)\Delta z_s$ into which a transverse distribution of microparticles will be assigned, each of which has a $z-$value at the centre of the slice. 
Note that the histogram bin width $\Delta z_h$, will usually be greater than the microparticle slice width $\Delta z_s$. Also, the slice width $\Delta z_s$ should  be sufficiently short to allow for the required resolution of microparticles in $z$.
For example in a FEL the initial microparticle spacing, $\Delta z_s$, should be significantly smaller than the resonant radiation wavelength $\lambda_r$ to allow the coherent bunching of the electrons at the radiation wavelength scale to be resolved. The number of slices per radiation wavelength should therefore be $N_\lambda \gg 1$ so that $\Delta z_s=\lambda_r/N_\lambda$. 

A transverse charge density function $f_\perp(x,y,z)$ is then created for each longitudinal slice using a similar 2-D histogram-interpolation method. Here the $z$-dependence of $f_\perp(x,y,z)$ will be in integer units of the slice width $\Delta z_s$. The longitudinal density function and the  transverse density function can then be interpolated to create a continuous 3-D charge density function, $f(x,y,z)=f_z(z)f_\perp(x,y,z)$.

Smoothing of the charge  distribution function may optionally be applied via a suitable smoothing function when applying the interpolation from the histogram.


As described in the next section, a new set of microparticles is then created in each longitudinal slice in $z$ via a 2-D Joint Distribution Function (JDF)~\cite{jdf,cdf} in the transverse plane so that the JDF $\propto f_\perp(x,y)$. The effects of Poissonian shot-noise are then added to each microparticle using the method of~\cite{shotnoise}. Note that the JDF retains greater information on the distribution of particles in the beam than by projection of charge first onto the transverse $x$ and $y$ axes independently to give the product of two 1D `Cumulative Distribution Functions' (CDF) so that  $f_\perp(x,y) \rightarrow  f_x(x)\times f_y(y)$. The microparticle distributions generated by the JDF method can then model more complex electron beam phase spaces.

\begin{figure}[hbt!]
\includegraphics[width=1.0\linewidth]{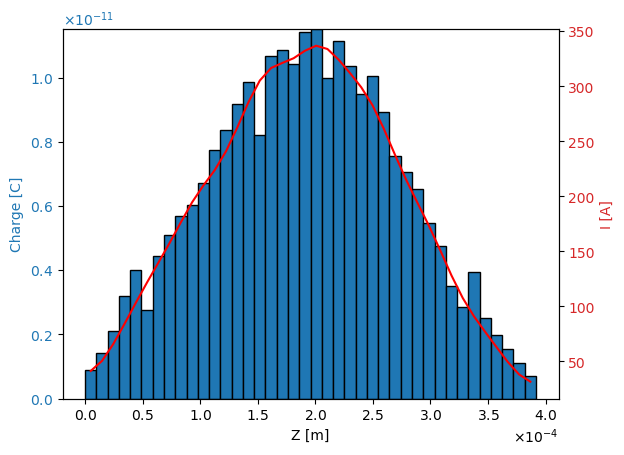}
\caption{Sample electron charge per bin along the $z$--axis. The plot shows the original binned histogram of the macroparticle charge and the interpolated data. The original beam is sampled over 50 bins and then smoothed using the {\it Python SciPy scipy.ndimage.gaussian} filter~\cite{PythonLibs} to give the (red) plot of the longitudinal current profile. The Gaussian smoothing filter parameter in this figure was  $\sigma_z=1.5\Delta z_h$.}
\label{El_Dens_Profile}
\end{figure}

The longitudinal charge histogram  and the corresponding current evaluated from the charge distribution function, $I(z)\approx cf_z(z)$ for the relativistic beams assumed here, is shown in the example of Fig.~\ref{El_Dens_Profile}. 
The electron beam  parameters  are similar to a beam generated in designs for the CLARA FEL test facility~\cite{CLARA} as shown in Table~\ref{table:1} which gives: the peak current; normalised RMS emittance; mean Lorentz factor; Lorentz factor spread; pulse duration; and electron bunch charge. For a typical undulator of period $\lambda_u\approx 3$cm this gives a FEL wavelength tuning of $\lambda_r \approx 100-400$nm.  
\begin{table}[h!]
\centering
\begin{tabular}{|c|c|c|c|c|c|c|} 
 \hline
 $I_{pk}$ & $\epsilon_{n}$[mm-mRad] & $\bar{\gamma}$ & $\sigma_\gamma$ & $\sigma_{z}$ [$\mu m$] & $Q$ [pC]\\ 
 \hline
 $395$ & $0.3$ & $475$ & $0.04$ & $82$ & $250$\\ 

 \hline
\end{tabular}
\caption{Typical CLARA beam parameters.}
\label{table:1}
\end{table}

Here a Gaussian smoothing function, with smoothing parameter of $1.5$ of the histogram bin width, $\sigma_z=1.5 \Delta z_h$ was used~\cite{PythonLibs}. Such smoothing can be important in FEL modelling as un-physical, sharp changes in current can lead to the spurious generation of significant Coherent Spontaneous Emission which would not be present in the real electron beam and could subsequently be amplified in the FEL simulation~\cite{shotnoise}. 

An optionally smoothed charge density distribution function in the transverse plane $f_\perp(x,y,z)$ is also calculated from the macroparticle distribution for each longitudinal slice in a similar way. 
This transverse charge density will, in general, be a function of the longitudinal  $z$ position of a given slice within the beam.
The result is an optionally smoothed, 3-D charge density distribution function of the macroparticle distribution, $f(x,y,z)=f_z(z)f_\perp(x,y,z)$. This function is significantly smoother than a discrete charge distribution of the initial, often relatively sparse, macroparticle distribution and allows for a less noisy, more realistic, distribution of microparticles to be generated for input into the FEL modelling software. 
The three dimensional density map (3D histogram) smoothing is obtained by using the Python built-in library {\it SciPy ndimage gaussian\_filter} function~\cite{PythonLibs}, which creates a convolution of the macroparticle histogram with a Gaussian function. The greater the width of the Gaussian function, the greater the smoothing, and the more the charge distribution function will deviate from the macroparticle charge distribution. On the other hand, an insufficient degree of smoothing will make the individual macroparticles more visible and can result in holes in the smoothed charge distribution function as a result of oversampling. A good starting point would be to choose the Gaussian width to be  approximately the mean macroparticle spacing.





\section{Microparticle generation \label{S3}}
The method as described in Section~\ref{S2} to generate the microparticles for use in the simulation of the FEL interaction is now demonstrated. 

\begin{figure}[hbt!]
\centering
\includegraphics[width=0.85\linewidth]{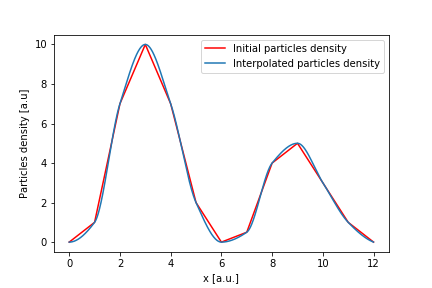}
\caption{The initial and smoothed transverse charge density function of the macroparticle distribution in the $x$--axis at a given slice in $z$. The beam centre here is at $x=6$. }
\label{dens_profile_example} 
\end{figure}

\begin{figure}[hbt!]
\centering
\includegraphics[width=0.85\linewidth]{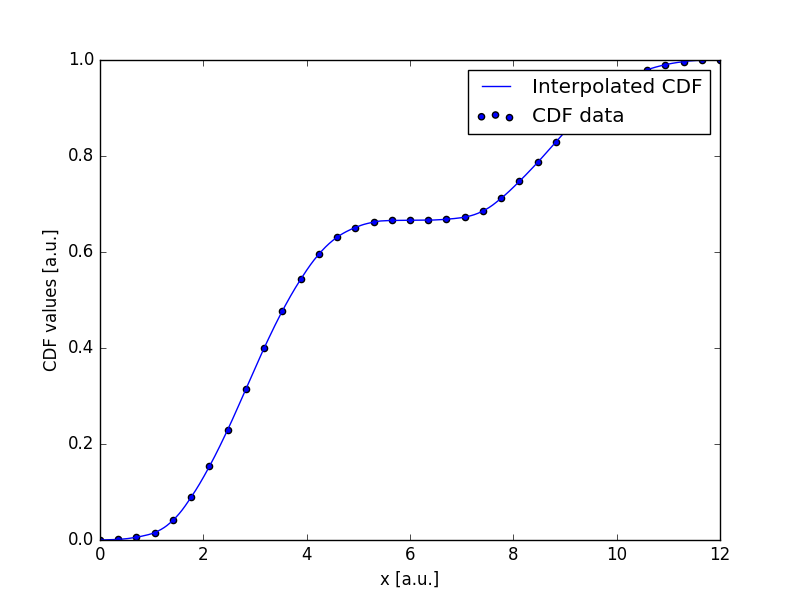}
\caption{The Cumulative Distribution Function  of a single slice taken from the macroparticle distribution Fig.~\ref{dens_profile_example} and its interpolated function.}
\label{cdf_line}
\end{figure}

The basic algorithm is first demonstrated for a simple case where only one transverse dimension, $x$, is used, i.e. via a CDF so that $f_\perp(x,y,z) \rightarrow f_\perp(x,z)$. For each beam slice along the $z$--axis, the initial macroparticle density is first smoothed to generate a transverse charge density for the slice as shown in the example of Fig.~\ref{dens_profile_example}.
The CDF is then calculated by integrating and normalising this density function as shown in Fig.~\ref{cdf_line}.

\begin{figure}[hbt!]
\centering
\includegraphics[width=0.85\linewidth]{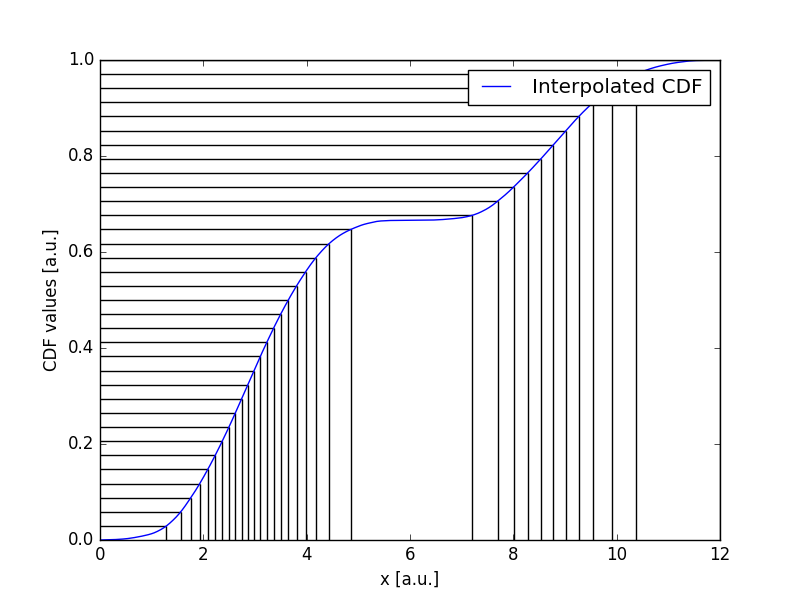}
\caption{The Cumulative Distribution Function profile of a single slice taken from the interpolated macroparticle distribution of Fig.~\ref{dens_profile_example}. On the vertical axis a series of equally spaced lines define the individual average microparticle charges. While these give microparticles of the same mean charge, they could also be randomly generated to give microparticles of different mean charge. The droplines from the CDF onto the horizontal axis then give the positions of the microparticles.}
\label{cdf_drop}
\end{figure}

The CDF is then used to generate the required number of microparticles re-creating the initial 1D transverse current density profile of the macroparticles, as shown in Fig.~\ref{cdf_drop}. An array of random values along the CDF vertical axis in the range $0$ to $1$, or equally spaced as shown here, is generated and the CDF is used to map these onto the microparticle positions along $x$ as shown. It is seen that the gradient of the CDF represents the particle density. Finally, the new microparticle distribution can be checked by creating a histogram and comparing it with that of the original macroparticle distribution histogram. This is repeated for each electron beam slice in $z$. 

\begin{figure}
\includegraphics[width=1.0\linewidth]{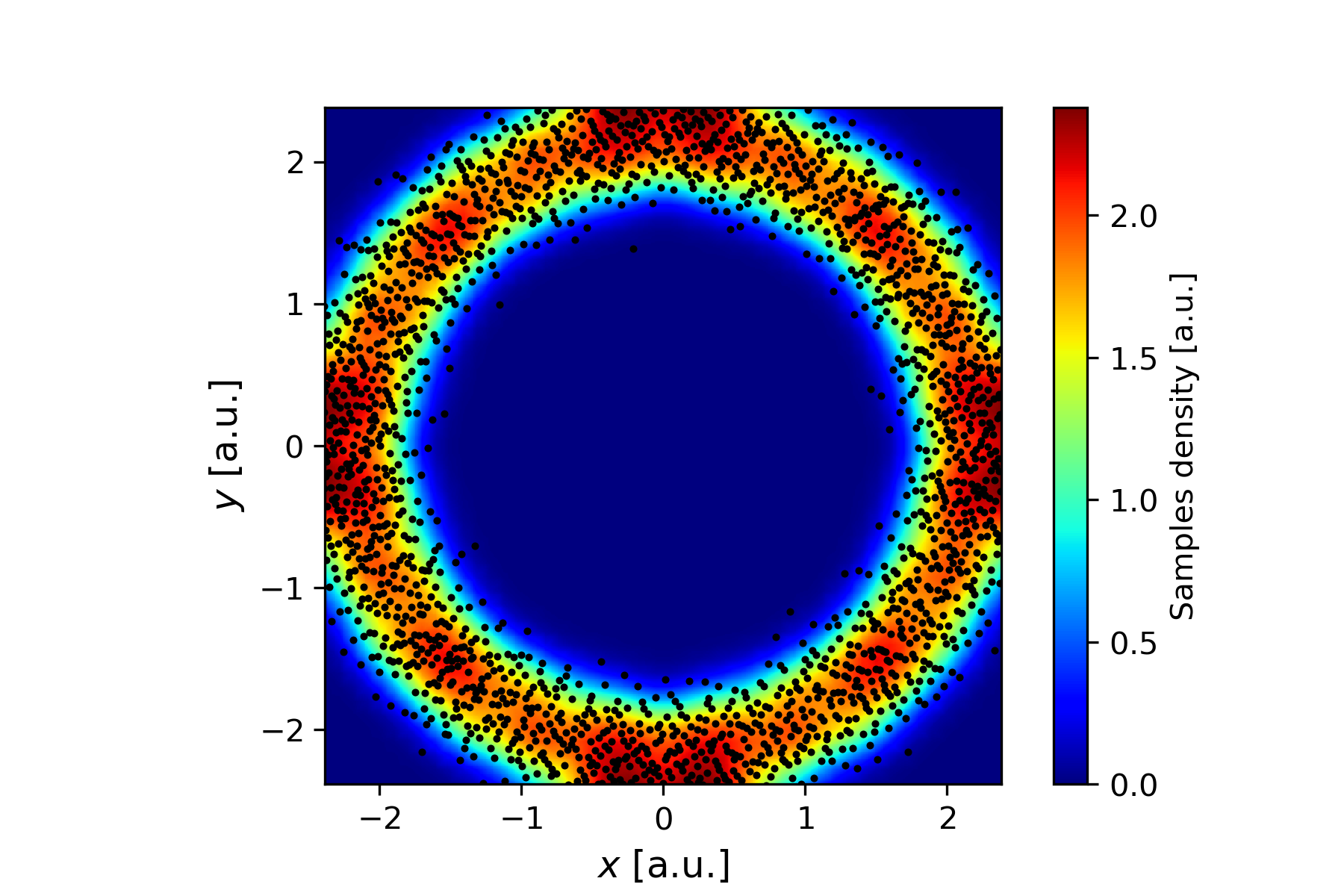}
\caption{An example of a electron density map of a slice of a hollow beam in the transverse ($x,y$) plane. The initial 150 macroparticles (not shown) were first binned over a 50$\times$50 histogram grid. A Gaussian smoothing, using a factor of 1.5 times this grid spacing, was the used to obtain the transverse charge density function, here plotted over 500$\times$500 points in a colour density map. The black dots are 2000 microparticles created using the JDF method.}
\label{jdf_circle}
\end{figure}

Note that the CDF function used to  demonstrate the process above is replaced in the developed software by a (parallelised) 2-dimensional JDF.
An example of a transverse structure that requires the use of a JDF is a `hollow' electron beam in the transverse plane. Fig.~\ref{jdf_circle} demonstrates the use of the JDF in the transverse cross--section of such a beam and is seen to maintain the hollow beam structure.



\section{Microparticle Momentum \label{S4}}
After the spatial generation of the microparticles from the macroparticle distribution as described above, the momenta of the macroparicles is mapped onto them. The initial 6-D phase space coordinates of the macroparticles creates a multidimensional map for the momentum which is interpolated, using their spatial coordinates, onto the momenta of the microparticles created in Section~\ref{S3}. 

This is performed using the {\it griddata} module of the Python {\it SciPy} {\it interpolate} library~\cite{NMSci,PythonLibs}. The {\it griddata} module interpolates the macroparticle momentum onto the microparticles using either nearest--neighbour or a spatial linear interpolarion approximation applied to the whole macroparticle distribution in one process, i.e. it is not applied on a slice-by-slice basis. The general principle underlying {\it griddata} is to tessellate the input data set to $n$-dimensional simplices, and then interpolate linearly on each simplex. When using the linear interpolation option, the interpolant is built by triangulating the input macroparticle data with the {\it Qhull} library~\cite{Qhull,Qhull2}, and then on each triangle performing a linear barycentric interpolation~\cite{NumRec} of the macroparticle momenta onto the microparticles. 

Note that the {\it griddata} module does not require a uniform grid and therefore works efficiently even with with relatively irregularly spaced macroparticles and microparticles.

Where there may be significant and potentially important correlations between some macroparticle momentum variables, e.g. between transverse momenta $p_x$ and $p_y$, it would be prudent to apply both nearest-neighbour and linear interpolation. One can then compare any important correlations in the final microparticle distributions. This would also allow any reductions in the spread in momenta in the microparticles, due to linear interpolation (regression towards the mean), to become apparent. This may then require a greater number of macroparticles for sampling via linear interpolation, or the use of the nearest-neighbour only sampling.  

\section{Poissonian noise application \label{S5}}
Following the assignment of the interpolated positions and momenta of the microparticles as described in Sections~\ref{S2}~--~\ref{S4},  a Poissonian noise is then applied to the microparticles to model the shot-noise of a relativistic electron beam. This is done by adding a Poisonnian noise contribution to both the interpolated mean microparticle positions in $z$, and to their charge weighting as detailed in~\cite{shotnoise}. The noise added to the microparticle distribution is compared with the theoretical shot-noise of a real relativistic electron beam propagating along the $z-$axis. 

Note that the Poissonian noise  added to the microparticles is independent of any FEL operating wavelength and can be regarded as a generic property of the relativistic beam shot-noise. 

Following generation of the  microparticle ensemble from the macroparticles as described above, they are equally spaced at a distance of $\Delta z_s=\lambda_r/N_\lambda$ along the $z$--axis in a sliced structure. 
A random variation about the mean $z$ position of each microparticle is then introduced by adding  $\left ( \Delta z_s/\sqrt{\bar{N}_e} \right ) R$ onto its mean position in $z$, where $R$ is a uniform random number $-0.5<R<0.5$, and $\bar{N}_e$ is the mean number of electrons (the mean charge weight) each microparticle represents~\cite{shotnoise}. The charge weight of each microparticle then also has noise added to it by assigning it a random charge weight $N_e$, which is generated by a Poisson random deviate generator of mean $\bar{N_e}$~\cite{shotnoise}. Note that the random charge weight $N_e$ is an integer, so that each microparticle will represent an integer number of electrons. For low charge weights, sometimes the Poisson random deviate generator may assign $N_e=0$. In this case these are simply removed from the microparticle distribution. 

\begin{figure}
\centering
\begin{subfigure}[b]{0.85\textwidth}
   \includegraphics[width=1.0\linewidth]{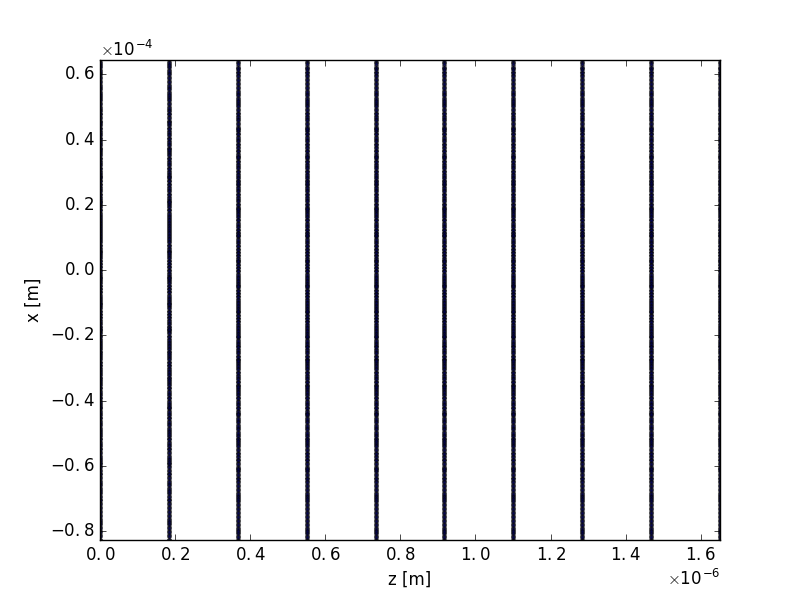}
   \caption{Microparticles without noise applied.}
   \label{fig:sub1} 
\end{subfigure}
\begin{subfigure}[b]{0.85\textwidth}
   \includegraphics[width=1.0\linewidth]{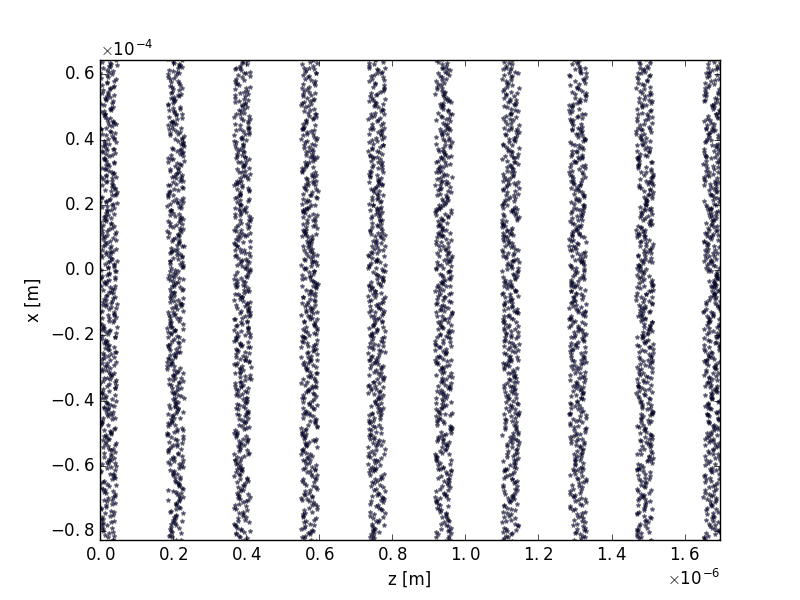}
   \caption{Microparticles with noise applied.}
   \label{fig1:sub2}
\end{subfigure}
\caption{The plots show $x/z$ projection of the beam before (top) and after applying the microparticle Poissonian noise (bottom). On both plots one can clearly observe that the beam has an ordered structure with microparticles equally separated along the longitudinal $z$--axis with a slice spacing $\Delta z_s\approx 183$~nm. For the purpose of the visualisation the noise has been amplified by factor of $5$.}
\label{fig1:noiseplot}
\end{figure}

Fig.~\ref{fig1:noiseplot} shows an example of a microparticle distribution along the longitudinal $z$-axis before and after adding shot-noise, in this case for slices of width of $\Delta z_s\approx 183$~nm. Before adding noise all microparticles in the same slice have equal values of $z$ (top), while after the Poissonian noise has been added, they spread about their initial mean positions (bottom). The microparticles also have the Poissonian charge variation applied as described above (not shown).

Coherent FEL radiation is driven by electron bunching at the radiation wavelength. At a given point in the electron beam a `bunching parameter' $b(z,t)=1/N\sum_j \exp(-i\theta_j)$ may be defined, with, $\theta_j=(k_r+k_u)z_j-\omega_rt$ and where $k_r$ and $k_u$ are the radiation and undulator wavenumbers respectively with $\omega_r=ck_r$.  Subscripts $j=1..N$ refer to the $N$ electrons contained within an interval $2\pi$ of $\theta$, approximately one radiation wavelength, about position $z$ where $b$ is calculated. Note that the shot-noise added to the microparticles depends only upon the local electron density and not the specific FEL parameters $k_r$ and $k_u$. The bunching parameter will therefore be correctly modelled for a wide range of wavelengths including harmonics of the resonant wavelength. 

\begin{figure}   
\label{sub1} 
\centering
\begin{subfigure}[b]{0.80\textwidth}
   \includegraphics[width=1.0\linewidth]{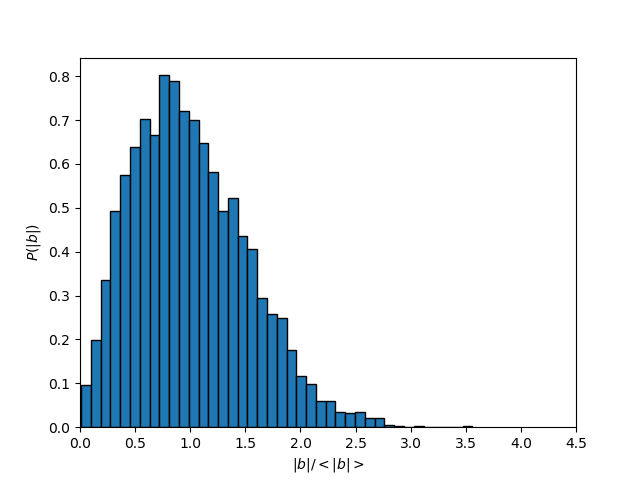}
\end{subfigure}
\begin{subfigure}[b]{0.80\textwidth}
   \includegraphics[width=1.0\linewidth]{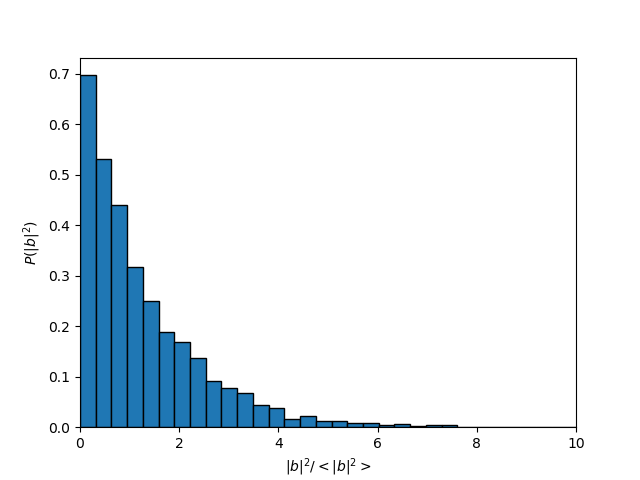}
\end{subfigure}
\caption{Microparticles bunching statistics after CDF is applied. Top: the probability distribution for the magnitude of the bunching parameter $|b|$ (Rayleigh distribution function) and bottom: the probability distribution for the magnitude of the bunching parameter squared $|b|^2$ (negative exponential distribution function). Both are in good agreement with theory~\cite{shotnoise}. }
\label{bunchstats}
\end{figure}

The results obtained by adding the shot-noise to the microparticles were tested by analysing an ensemble of identical microparticles with different noise added. Each member of the ensemble would give the initial conditions for an electron beam before propagating through an FEL. 
The scaled probability distributions for the bunching magnitude $|b|$ and its square $|b^2|$ are shown in Fig.~\ref{bunchstats}, and agree very well with the Rayleigh and exponential distribution functions, expected from the previous work of~\cite{shotnoise}. While the parameters used here were for an FEL operating with wavelength $\lambda=4 \mu$m, undulator period $\lambda_u=4$cm and $\bar{\gamma}=100$, the results obtained in Fig.~\ref{bunchstats} are generally valid for an electron beam with shot-noise obeying Poissonian statistics.

\begin{figure}
\includegraphics[width=1.0\linewidth]{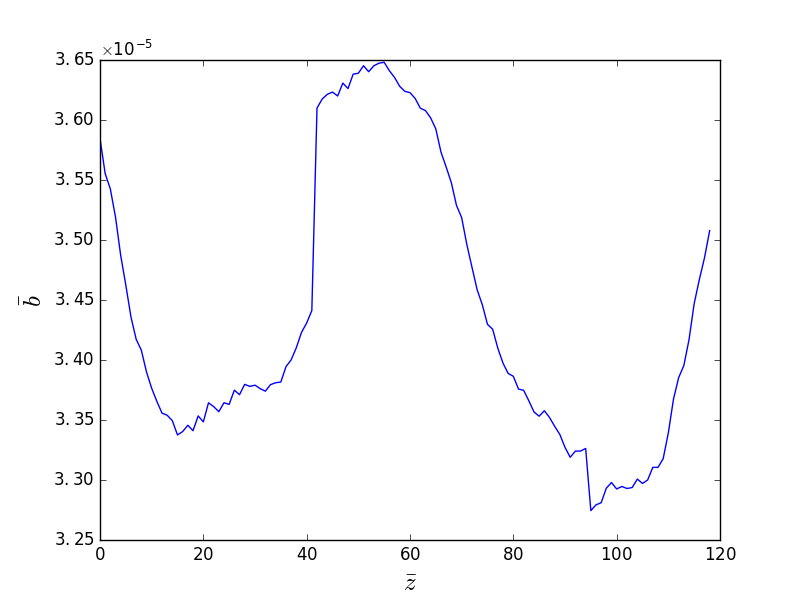}
\caption{Bunching in an electron beam propagating through the undulator with FEL interaction turned off. The mean value of $|b|$ is predicted to be  $\bar{b} \approx 3.97\times 10^{-5}$ and its standard deviation $\sigma_{|b|}$ is $2.077\times 10^{-5}$. The numerical results are seen to be  within the limits as predicted be the theory. Here $\bar{z}=z/l_g$ is the distance along the undulator axis scaled with respect to the FEL gain length~\cite{np}}
\label{bunch_sweep}
\end{figure}

An additional test was performed to verify how the bunching behaves as the microparticles propagate through an undulator, but in the absence of any FEL interaction. When FEL action is switched off, the bunching parameter can be expected to evolve as the microparticles move from their initial positions due to emittance and energy spread effects.  The bunching parameter $|b|$ at a given point should therefore evolve as the beam propagates through the undulator.  The FEL simulation software Puffin~\cite{puffin,puffin2} was used to propagate the beam without any FEL interaction along an undulator axis $\bar{z}$. The results are shown in Fig.~\ref{bunch_sweep} and the bunching $|b|$ is seen to evolve about its mean theoretical value and within the range $\sim \sigma_{|b|}$ in good agreement with the theoretical model of~\cite{shotnoise}.

\section{Comparison between input and output data}
An example of the conversion of an electron bunch of macroparticles, generated by the accelerator modelling code ASTRA~\cite{PAC2003, ASTRA} for a CLARA FEL~\cite{CLARA} simulation, into an electron bunch of many more microparticles suitable for the FEL simulation is now presented. Beam parameters are similar to those of Table~\ref{table:1}.

\begin{figure}
\includegraphics[width=1.0\linewidth]{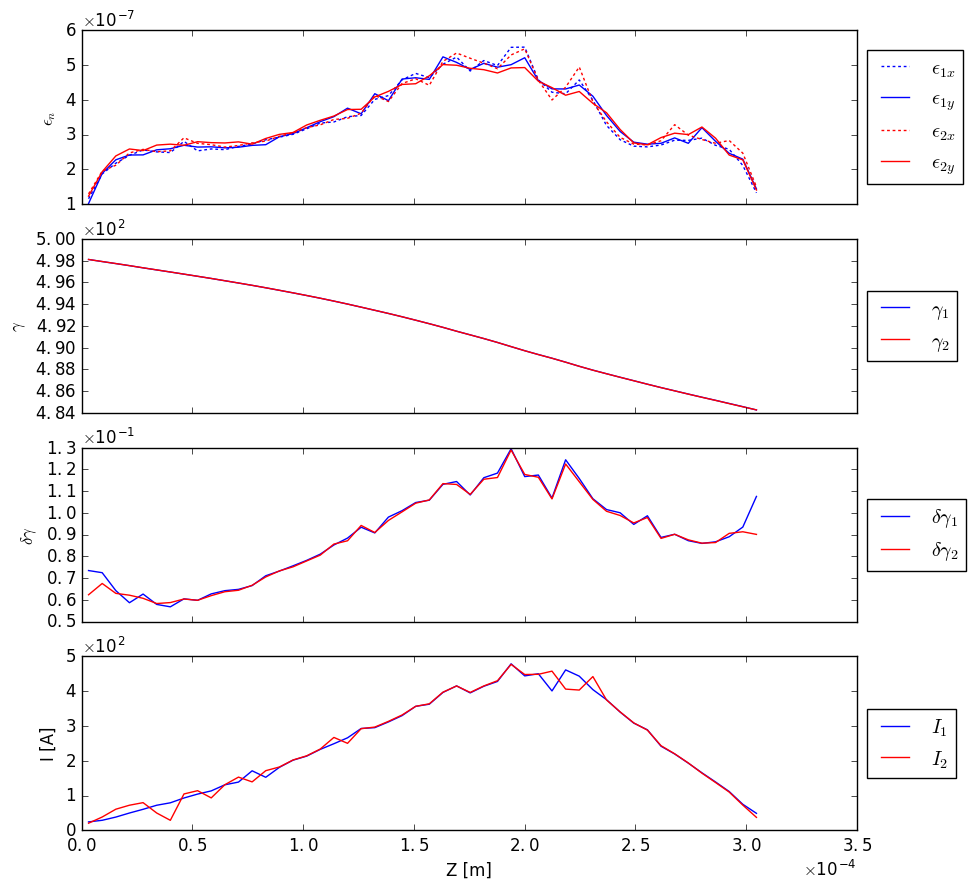}
\caption{Electron beam parameters as a function of position in the beam propagation direction $z$. From top: normalised emittance (m rad); beam lorentz factor ($\propto$ beam energy); energy spread and current as calculated from the macroparticle beam (blue) and microparticle beam after the JDF method has been applied (red).}
\label{fig3:beamtats}
\end{figure}

In changing from the macroparticle to microparticle distribution, beam parameters, such as current, energy spread, emittance etc, should remain essentially unchanged. This can be seen from  Fig.~\ref{fig3:beamtats} which plots the electron pulse emittance, energy, energy spread and current, as calculated from both the initial macroparticle distribution from the output from ASTRA with approximately $2.6\times 10^5$ macroparicles, and following application of the JDF microparticle generation of approximately $5.7\times 10^7$ microparticles with addition of shot-noise as described above. The final number of microparticles is the result of generating 20 slices per wavelength (71325 slices in total) and with each slice containing 800 microparticles. This microparticle distribution can then be used as input to the FEL simulation code Puffin. Note that, as expected, the plots obtained for the microparticle distribution are a little smoother than the original macroparticle distribution and is due to the increased microparticle density over that of the macroparticles. Further smoothing can also be applied as described above.


\begin{figure*}
\centering
\begin{subfigure}{.5\textwidth}
  \centering
  \includegraphics[width=1.0\linewidth]{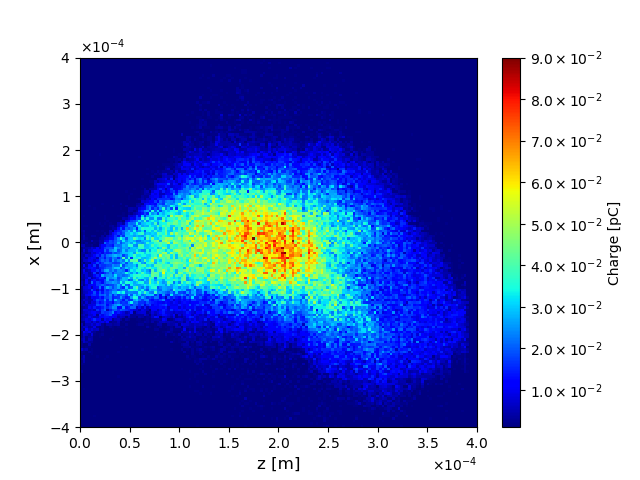}
  \label{fig5:sub3}
\end{subfigure}%
\begin{subfigure}{.5\textwidth}
  \centering
  \includegraphics[width=1.0\linewidth]{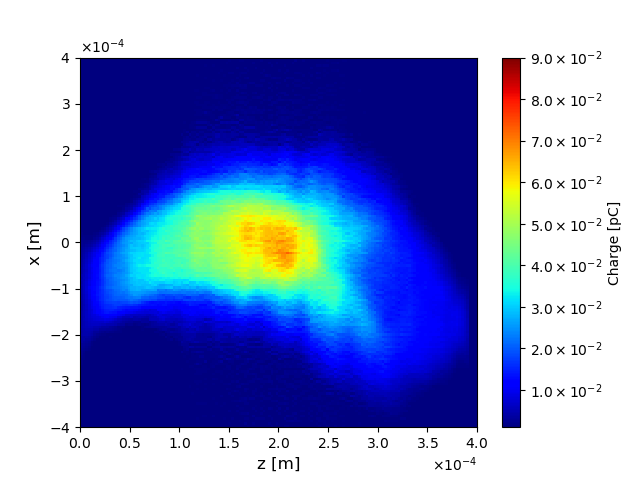}
  \label{fig5:sub4}
\end{subfigure}
\begin{subfigure}{.5\textwidth}
  \centering
  \includegraphics[width=1.0\linewidth]{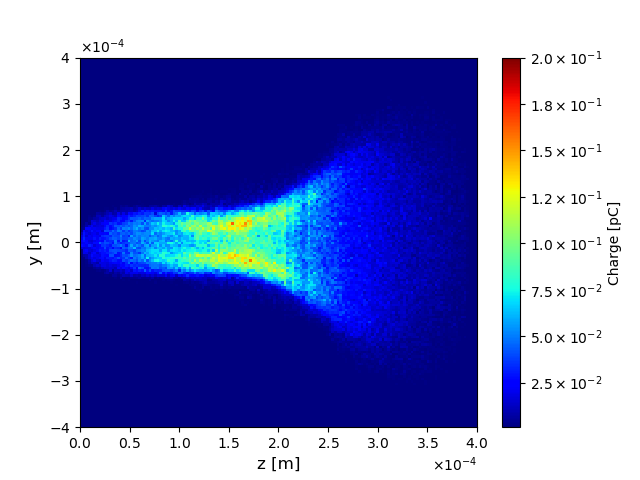}
  \label{fig6:sub1}
\end{subfigure}%
\begin{subfigure}{.5\textwidth}
  \centering
  \includegraphics[width=1.0\linewidth]{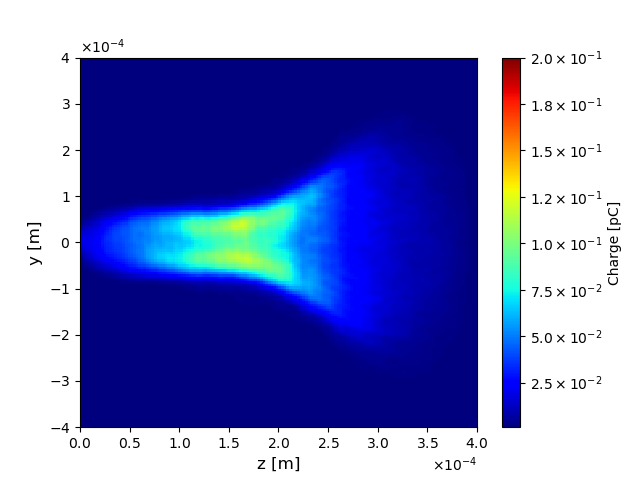}
  \label{fig6:sub2}
\end{subfigure}
\begin{subfigure}{.5\textwidth}
  \centering
  \includegraphics[width=1.0\linewidth]{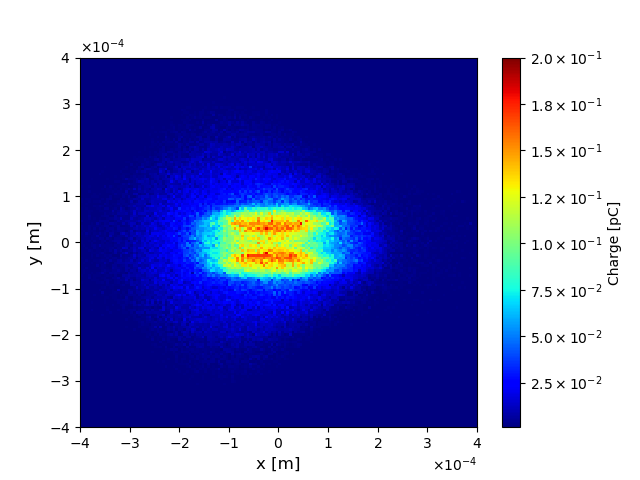}
  \label{fig6:sub3}
\end{subfigure}%
\begin{subfigure}{.5\textwidth}
  \centering
  \includegraphics[width=1.0\linewidth]{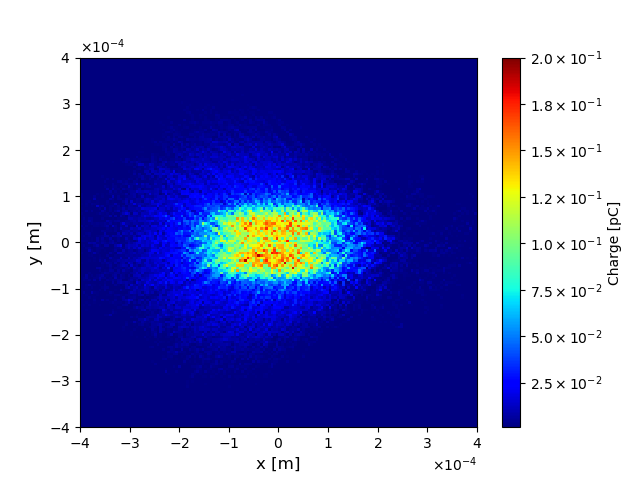}
  \label{fig6:sub4}
\end{subfigure}
\caption{Particle spatial density maps for macroparticles (left) and the microparticle distributions after JDF processing (right). The data was sampled over 150$\times$150 grid.}
\label{fig10:Density1}
\end{figure*}

\begin{figure*}
\centering
\begin{subfigure}{.5\textwidth}
  \centering
  \includegraphics[width=1.0\linewidth]{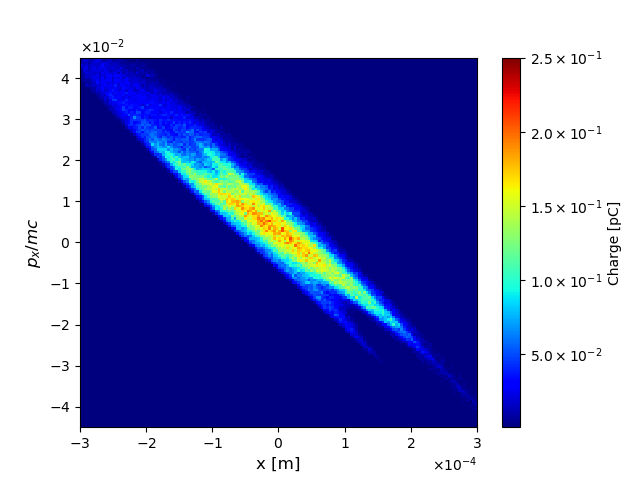}
  \label{fig5:sub1}
\end{subfigure}%
\begin{subfigure}{.5\textwidth}
  \centering
  \includegraphics[width=1.0\linewidth]{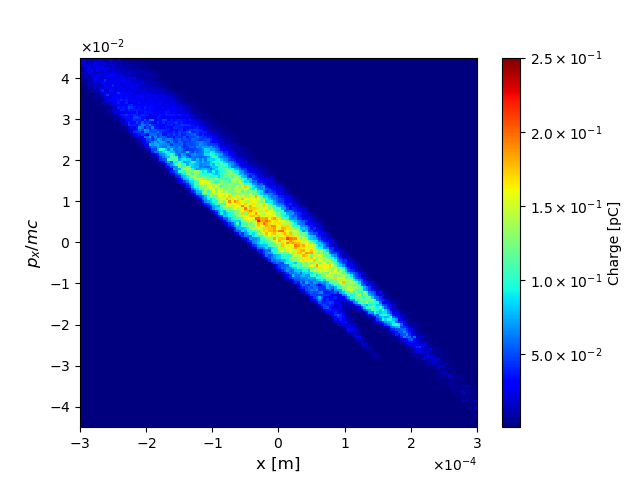}
  \label{fig5:sub2}
\end{subfigure}
\begin{subfigure}{.5\textwidth}
  \centering
  \includegraphics[width=1.0\linewidth]{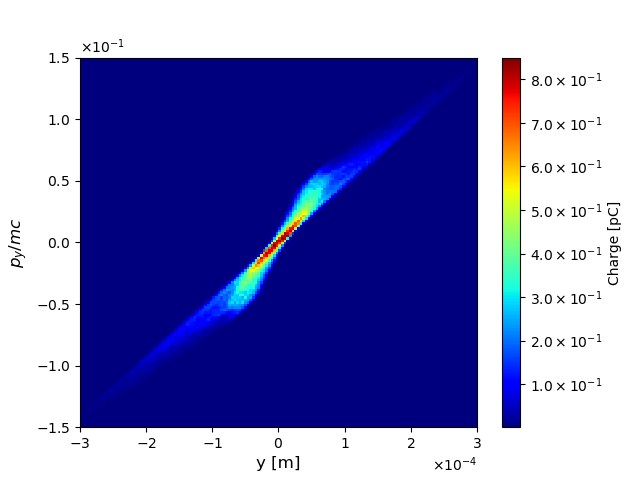}
  \label{fig4:sub3}
\end{subfigure}%
\begin{subfigure}{.5\textwidth}
  \centering
  \includegraphics[width=1.0\linewidth]{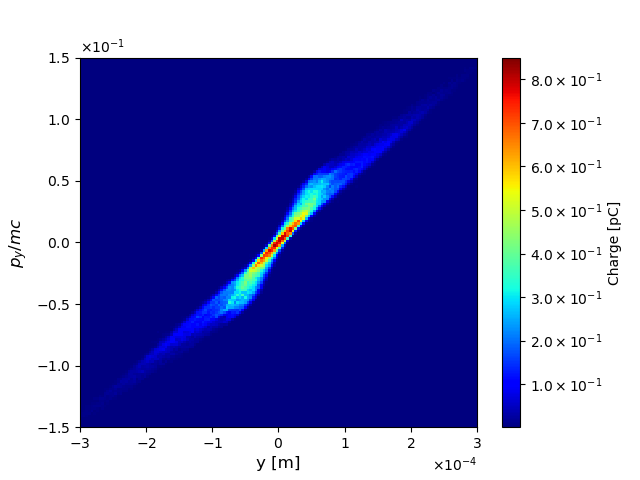}
  \label{fig4:sub4}
\end{subfigure}
\begin{subfigure}{.5\textwidth}
  \centering
  \includegraphics[width=1.0\linewidth]{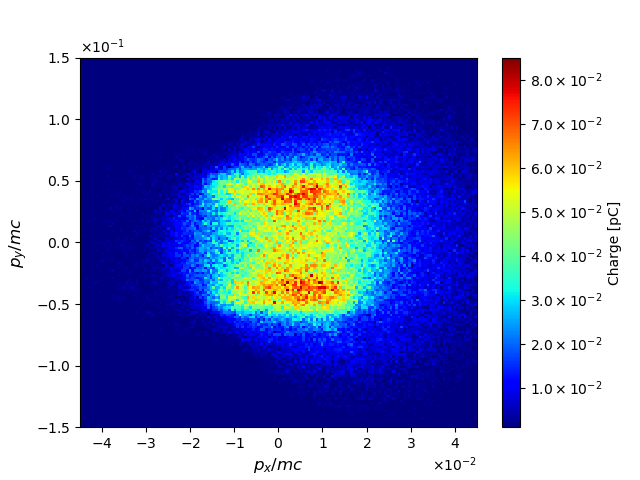}
  \label{fig4:sub1}
\end{subfigure}%
\begin{subfigure}{.5\textwidth}
  \centering
  \includegraphics[width=1.0\linewidth]{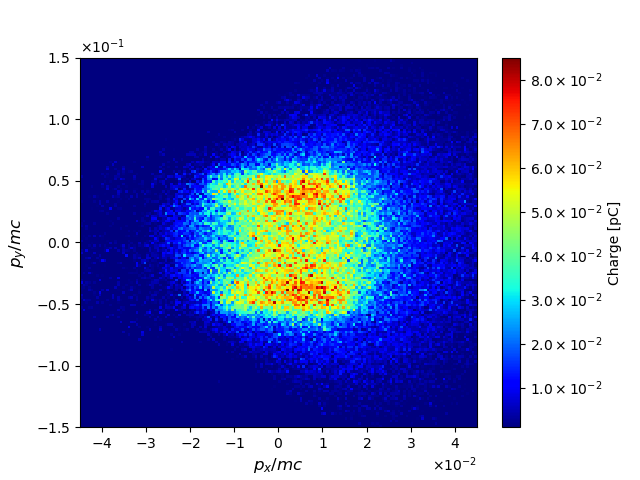}
  \label{fig4:sub2}
\end{subfigure}
\caption{Particle momentum density maps for macroparticles (left) and the microparticle distributions after JDF processing (right). The data was sampled over 150$\times$150 grid.}
\label{fig11:Density1}
\end{figure*}

In addition to the longitudinal beam parameters of Fig.~\ref{fig3:beamtats}, which are seen to be in in good agreement, further analysis was done on  projections of spatial and momentum density maps. These are shown for the spatial dimensions in Fig.~\ref{fig10:Density1} and for the momentum in Fig~\ref{fig11:Density1}. Again good agreement is seen. 

\section{Conclusion}
The method presented of increasing the number of simulation particles that describe a particle beam, while retaining a good estimate of the beam parameters and introducing realistic Poissonian shot-noise properties, has been described and demonstrated. 
While the method is primarily intended as an up-sampling of particles for subsequent injection into a FEL simulation code, it may well have applications in other areas, or be adapted to them, where the statistical noise may be important. 
The method is computationally efficient and easy to apply in most  computer programming languages. We note that the algorithm was previously successfully applied in so-called start--to--end simulations of the CLARA FEL using Puffin FEL simulation  software, and the FEL output data compared with same simulation using a different 3-D FEL simulator GENESIS~1.3~\cite{IPAC18,genesis}. Very good agreement between the two were found. The main limitation for the algorithm is in the potential sparcity of the input macroparticle ensemble. If it is too sparse the interpolation and extrapolation parts of the algorithm may create inaccurate results -- the quality of the output is only as good as that of the input. There is no easy way we have found to determine \emph{a-priori} how dense the input macroparticles should be to avoid potential problems - this probably needs to be done via trial and error. We would also advise use and comparison of both nearest-neighbour and linear interpolation methods in calculating the momentum of the microparticles from the macroparticles. The software and documentation is freely available from the repository~\cite{JDF_github}. 

\section{Acknowledgements}
We are grateful to funding from the Science and Technology Facilities
Council (Agreement Number 4163192 Release \#3); EPSRC grant EP/K000586/1; EPSRC
Grant EP/M011607/1

\end{document}